\documentclass[%
 aip,
 amsmath,amssymb,
 reprint,%
]{revtex4-1}

\usepackage{graphicx}
\usepackage{dcolumn}
\usepackage{bm}

\usepackage[utf8]{inputenc}
\usepackage[T1]{fontenc}
\usepackage{textcomp}
\usepackage{mathptmx}
\usepackage{bibentry}
\usepackage{subfig}

\DeclareMathAlphabet{\mathcal}{OMS}{cmsy}{m}{n}

\begin{document}

\preprint{AIP/123-QED}

\title[The regularized Stokeslets method applied to the three-sphere swimmer model.]{The regularized Stokeslets method applied to the three-sphere swimmer model.}

\author{Henrique N. Lengler}
 \email{henrique.lengler@ufrgs.br}
\affiliation{
              Instituto de Física\\
	      Universidade Federal do Rio Grande do Sul\\
	      Porto Alegre, 91501-970, Brazil.\\
}

\date{\today}

\begin{abstract}
We investigate the applicability of the Method of Regularized
Stokeslets (MRS) in the simulation of micro-swimmers at low Reynolds number.
The
chosen model for the study is the well-known three linked spheres swimmer.
We compare our results  
with the lattice Boltzmann method, multiparticle
collision dynamics, a numerical solution of the Oseen tensor equation and an analytical solution,
all taken from Earl \textit{et al.} [J. Chem. Phys. 126, 064703 (2007)].
The MRS is studied in detail, and our results show an excellent agreement with the
lattice Boltzmann method, and with the analytical solution in its range of validity.
We conclude that the MRS is well suited for this type of simulation, offering
advantages such as being easy to implement and to represent complex
geometries. Therefore it presents
itself as a suitable candidate for more complex simulations.
\end{abstract}

\maketitle

\section{\label{sec:level1}Introduction} 
The interest in the study and development of microswimmers has been growing in
the past years.\\
Microswimmers are mechanisms, whether biological or not, of 
microscopic dimensions that propels itself in a fluid. Some examples are biological
creatures like bacteria and human-made micro-robots. The study of the individual
and collective behaviour of these small machines has led to the discovery of
many new and curious phenomena, and they are currently objects of interest in many 
lines of research \cite{Roadmap}.\par
The locomotion and interaction of
microscopic swimmers in newtonian and incompressible fluids can be studied using
the mechanical equations. At such small scales and low velocities, the Reynolds number is small,
and a simplified linear
approximation of the Navier-Stokes equations can be used \cite{Childress}. The linear Stokes
equations, as it is called, is obtained by disregarding the inertial terms, given
the dominance of the viscous force at this scale.
In this process, we
remove any non-linear term, and also any time dependence from the equations.\par
The inexistence of time reflects the fact that at this regime, fluids respond
instantly to perturbations, and they dictate the time evolution of the
physical quantities of the fluid. As a consequence, if a force suddenly stops
acting on the fluid, the generated flow also vanishes suddenly.  Additionally,
any time external forces are inverted, an inverted flow pattern takes place.
These and other properties make low Reynolds number flows unique, and are
responsible for some very curious phenomena, such as the possibility to
reverse fluid mixing under certain circumstances \cite{Unmixing}. They also impose a set of
conditions for autonomous swimming.\par
As explained by Purcell in his famous
paper \cite{Purcell}, only mechanisms that execute a
nonreciprocal sequence of movements, that is, movements that do not look the
same when analyzed backwards in time, are capable of travelling arbitrary long
distances in such environments.
One of the simplest swimmers that satisfies these conditions is the three-sphere
swimmer proposed in 2004 by Najafi and
Golestanian \cite{Najafi} and further analyzed in 2008 \cite{Najafi2}. Since then, this
model has been extensively studied by numerical, analytical and experimental methods \cite{Effic1,
Experimental,PhysRevE.85.061914}. Because of its simplicity and the possibility of
analytical studies, it
can serve as a good initial test for numerical methods that may later be used
for more complex systems (although there is another simple model \cite{Avron_2005} that
could also be used).\par
The study of such mechanisms by means of the linear
equations is not always trivial. The linearization  is generally not enough to
make the task of predicting fluid behaviour easy. Usually, only trivial cases
with simple geometries or few constituents can be studied analytically in great
detail.
For this reason, there is still interest in the development and study of
new methods for simulating low Reynolds number interactions. Nowadays, highly used
methods for these situations are the multi-particle
collision dynamics (MPC) and the lattice Boltzmann method (LBM). Both have very
different approaches, merits and limitations. The MPC and LBM methods, together with a
numerical solution of the Oseen tensor equations (OTE) and an analytical approximation, have been explained and compared in
the specific case of the three linked spheres swimmer in \cite{Yeomans}. Here, based on
this work, we proceeded to add a fourth method in the comparison, namely the
Method of Regularized Stokeslets (MRS) \cite{Cortez2d}. For this comparison, we 
implemented the MRS for the same system to compare to MPC, LBM, OTE and the analytical
approximation.
Our results show that the MRS
is well suited for this type of simulation, showing good agreement with the analytical
solution in the valid domain.
We finish by concluding that the MRS is a useful tool to be
used in the study of interactions at low Reynolds number.
We also discuss the peculiarities of the method and its numerical implementation details.

\section{The method of regularized Stokeslets}
\label{sec:the_method_of_regularized_stokeslets}
The MRS is based on a slight modification of
the Green function method for the linear Stokes equations. The Green function 
response for a delta distribution has a singularity at the perturbation point.
Therefore it is not much useful when used in discrete combinations,
since it adds singularities to the flow, not being very representative
of any physical behaviour.
It can be useful
in situations where the force of interaction on a continuous boundary is known
at each point, or a realistic one can be guessed.
In this case, it can be integrated to give
the total flow generated by this interaction.

\begin{figure}
	\centering
	\includegraphics[scale=0.70]{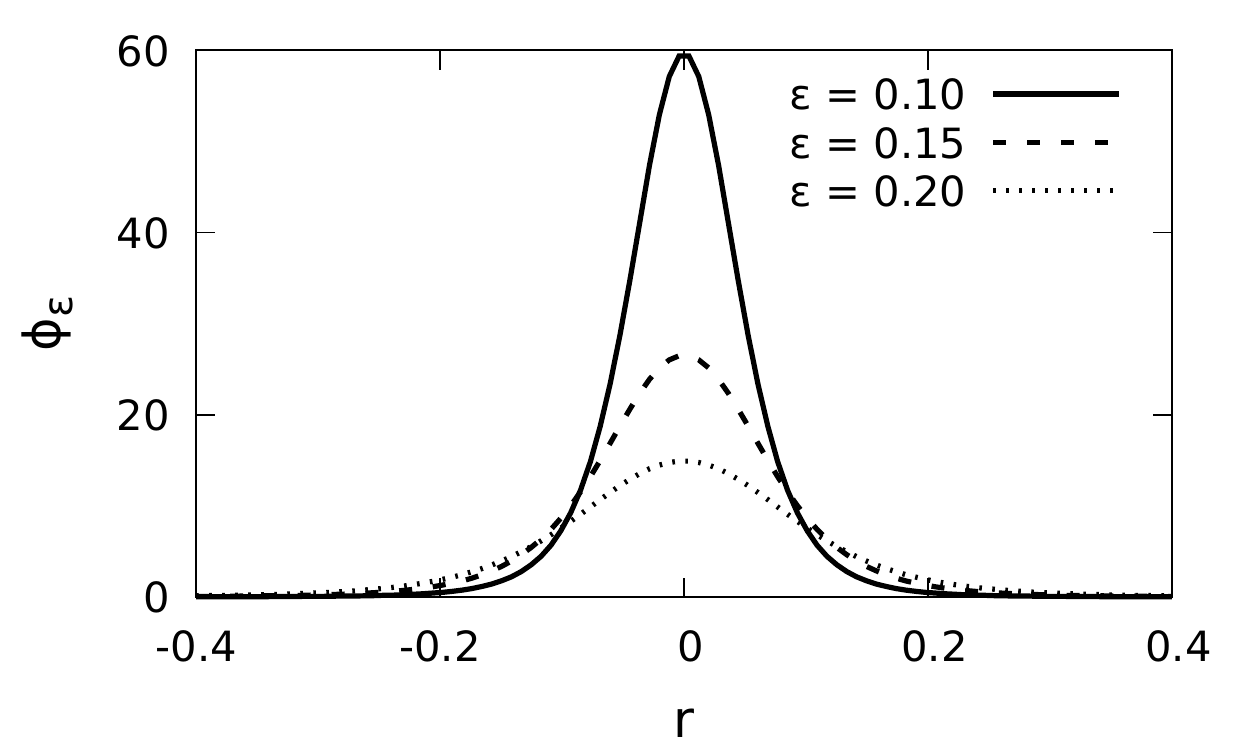}
	\caption{$\phi_{\epsilon}$ given in Eq. (\ref{eq:phi}) for the indicated parameters, a smaller $\epsilon$ results in a taller
	and more localized function. It approximates a Dirac delta in the limit $\epsilon \rightarrow 0$.}
	\label{fig:blob_example}
\end{figure}

In contrast with the standard Green
method, in the MRS, the delta distribution is replaced by a smooth,
radially symmetric and normalized function over the whole space. This function
is  controlled by a parameter $\epsilon > 0$ that determines how localized the
force is.\\ 
The equations to be solved are:
\begin{equation}
	\mu \nabla^2 \bm{u} = \nabla p - \bm{f}\phi_{\epsilon }
	\label{eq:geral1}
\end{equation}
\begin{equation}
	\nabla \cdot \bm{u} = 0
	\label{eq:geral2}
\end{equation}
Where $\bm{u}$ is the fluid velocity, $\mu$ is the
viscosity, $p$ is the pressure, $\bm{f}$ is a constant vector representing the
interaction force and
\begin{equation}
	\phi_{\epsilon} = f(\bm{x}) = g(\left| \bm{x} - \bm{x_0} \right|)
	\label{eq:phi}
\end{equation}
is the chosen regularized delta, dependent only on the distance from the
perturbation $ \bm{x}_0$.\\
In this paper, we use the amply used  $\phi_\epsilon(r)$ given by \cite{Cortez2d} and shown in Fig. \ref{fig:blob_example}.
\begin{equation}
	\phi_{\epsilon} (r) = \frac{15 \epsilon^4}{8 \pi (r^2 + \epsilon^2)^{7/2}} 
\end{equation}
Equations (\ref{eq:geral1}) and (\ref{eq:geral2}) are solved by:
\begin{equation}
	\mu \bm{u}(\bm{x}) = (\bm{f} \cdot \nabla )\nabla B_{\epsilon}(\bm{x} - \bm{x}_0) - \bm{f} G_{\epsilon}(\bm{x} - \bm{x_0})
\label{eq:general_solution}
\end{equation}
valid for the 2D and 3D cases (derived in \cite{Cortez2d} together with an
expression for the pressure). $G_{\epsilon}(r)$ and $B_{\epsilon}(r)$ are auxiliary functions defined
as solutions of $ \nabla^2 G _{\epsilon}(r) = \phi_{\epsilon} (r)$ and $ \nabla^2 B_{\epsilon}(r)= G _{\epsilon}(r)$, 
for $r = \left| \bm{x} - \bm{x}_0 \right|$ and,
in all equations the vector operators act on the cartesian coordinates
$\bm{x}$.

Interestingly, this type of perturbation generates a finite and
non-singular response at the point of perturbation, allowing the no-slip
condition to be imposed at $\bm{x} = \bm{x}_0$, leading to the possibility of using
these perturbations to represent small particles. The response now can be interpreted
as a velocity field generated by a mean interaction over a ball.
We can also use a finite, discrete and closely placed set
of such perturbations to represent a surface interaction.
Since the equations (1) and (2) are linear,
the velocity response of multiple perturbations can be constructed by a linear
combination. If we have $N$ interactions with the fluid, each one exerting a
force $\bm{f}_k$ at points $\bm{x}_k$, we can build the solution:
\begin{equation}
	\bm{u}(\bm{x}) = \bm{U}_0 + \frac{1}{\mu} \sum^{N}_{k=1} { ( \bm{f}_k
	\cdot \nabla ) \nabla B_{\epsilon} (r_k) - \bm{f}_k G_{\epsilon}(r_k) }
	\label{eq:solution2}
\end{equation}
for $r_k = \left| \bm{x} - \bm{x_k} \right|$. The expression within the summation can
be expanded and simplified given that $B$ and $G$ are dependent on
$\left| \bm{x} - \bm{x_k} \right|$ only, as also
shown in \cite{Cortez2d}. 
Given a choice of $\phi_{\epsilon}$ we can find both $G$ and $B$ by supposing $G$ and $B$ radially symmetric. Any constant of integration can be adjusted so that there is no flow for $r \rightarrow \infty$ (this is possible in three dimensions), and to make the velocity finite at each perturbation ($r=0$).
Any other constant term can be eliminated by the choice of $\bm{U}_0$, in our case we can set $\bm{U}_0 = \bm{0}$.

Eq. (\ref{eq:solution2}) can be used to compute flows if we know the forces
of interaction.  In general, we only know the velocities
of each point, and due to the regularization, the
no-slip condition can be imposed at each point
$\bm{x}_i$:
\begin{equation}
	\bm{u}(\bm{x}_i) = \frac{1}{\mu} \sum^{N}_{k=1} { ( \bm{f}_k
	\cdot \nabla ) \nabla B_{\epsilon} (r_{ik}) - \bm{f}_k G_{\epsilon}(r_{ik}) }
\label{eq:solution_sum}
\end{equation}
where $r_{ik} = \left| \bm{x}_i - \bm{x}_k \right|$. This sum can
be seen as:
\begin{equation}
	\bm{u}( \bm{x}_i) \equiv \bm{u}_i = \sum^{N}_{k=1} \bm{M(r_{ik})} \bm{f}_k
\end{equation}
where each term is composed of a linear operator $\bm{M}$ dependent on the distances,
acting on $\bm{f}_k$. If $D$ is the dimension, the operator $\bm{M}$ acts
on $\mathbb{R}^D$, and its matrix representation has size $D^2$. However,
the whole system can be seen as a linear system in $\mathbb{R}^{DN}$:
\begin{equation}
	\mathcal{U = MF}
	\label{eq:lin_system}
\end{equation}
if we treat $\mathcal{U}$ and $\mathcal{F}$ as augmented vectors of size $D \cdot N$
and $\mathcal{M}$ as the augmented matrix of all $\bm{M}$'s.
Since now we know the velocity of each particle, we can solve Eq. (\ref{eq:lin_system})
numerically for the forces and then return to Eq. (\ref{eq:solution2})
to compute the flow.
Generally, the matrix $\mathcal{M}$ is not invertible, but we can find solutions with
iterative methods. In this paper, we used GMRES with zero initial guess in every case.
With a choice of $\phi_{\epsilon}$ we can find the auxiliary functions, and then the
expressions for each operator $\bm{M}$ and consequently for $\mathcal{M}$.
Recalling $\phi_{\epsilon}$ from Eq. (\ref{eq:phi}), the expression for the operator is: 
\begin{equation}
	\left[\bm{M}(r_{ik})\right]_{lm} = \frac{1}{\mu} \{ F_1(r_{ik}) \delta_{lm} + F_2(r_{ik})(\bm{r}_{ik})_l(\bm{r}_{ik})_m\}
\end{equation}
with:
\begin{equation}
	F_1(r) = \frac{1}{8 \pi} \frac{r^2 + 2 \epsilon ^2}{(r^2 + \epsilon ^2)^{3/2}}\\
\end{equation}
\begin{equation}
F_2(r) = \frac{1}{8 \pi(r^2 + \epsilon ^2)^{3/2}} 
\end{equation}
\begin{figure}[h]
\centering
\includegraphics[]{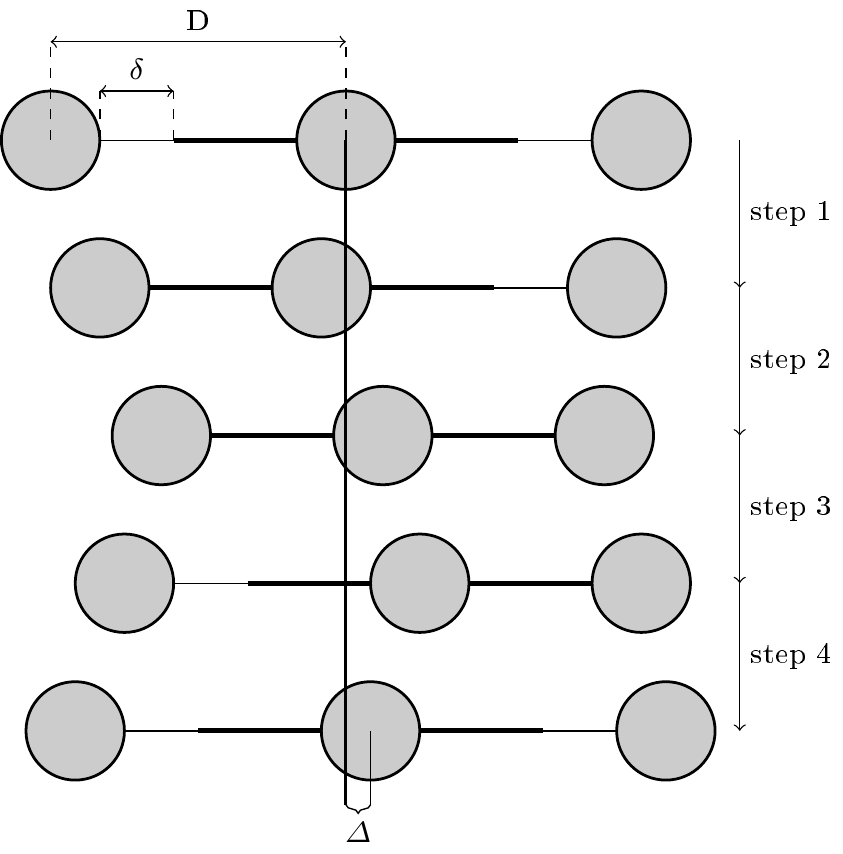}
\caption{Qualitative view of the swimmer motion in a complete cycle. After four steps it returns to its original
configuration, but in a different location.}
\label{fig:swimmer}
\end{figure}
\section{The three-sphere swimmer}%
\label{sec:the_three_sphere_swimmer}
For the comparison, we analysed the swimmer proposed in \cite{Najafi}
and studied by multiple methods in \cite{Yeomans}, the three-sphere swimmer.
This swimmer consists of three spheres of radius $R$, connected on a line by two arms of negligible 
thickness. The swimmer moves by changing its arm's lengths in a specific manner so that
the complete sequence is non-reciprocal. The complete cycle consists of four steps, wherein at each step, one arm is
kept fixed while the length of the other is changed by an amount we define as $\delta$ with a constant rate.
That is illustrated in Fig. \ref{fig:swimmer}.
After one complete cycle, the swimmer returns to its original configuration, and we measure
$\Delta$, the translated distance.
\section{Numerical Study}%
\label{sec:numerical_study}
\subsection{Validation and tests}%
\label{sub:validation_and_tests}
Before using the method for the swimmer, we decided to validate and test
our implementation with the case of a single sphere translating
with constant velocity, a case for which we had data to compare \cite{Cortez3d, Thompson}.
Also, since the swimmer consists of three translating spheres,
we used these tests to decide what values of $N$, $\epsilon$ and what type of 
discretization to use.

The choice of $N$ must be carefully taken since it is the parameter that 
has the biggest impact on computational time and memory usage. We recall
that the matrix $\mathcal{M}$ has $(ND)^2$ terms.
Ideally, for the best precision, $N$ should be set as big as possible, with
$\epsilon$ approaching zero. If $N$ is too small, the set of points will not
represent well the surface of the sphere, and we would get poor results.
Due to the limited memory and computing power, we must find a balance between
memory, speed and precision.
A drawback of the method is that the matrix $\mathcal{M}$, Eq. (\ref{eq:lin_system}), is generally
not sparse. Its sparsity depends on the configuration of the points.
Because of that, it is not possible to reduce memory usage by using alternative 
storage methods for sparse matrices. Luckily the MRS enables us to get 
good results by using the strategy of decreasing the number of points and increasing the
volume of interaction by increasing $\epsilon$, and in general, as in the case
of our simulations, memory requirements were easily achievable.

For every value of $N$, we have to adjust $\epsilon$.
There is no general rule to find the best value of $\epsilon$ for a given 
$N$ \cite{Jesus}.  In general, it depends upon the distances between points.
Our approach is to choose $\epsilon$ after defining $N$, by varying it until
we get enough precision. For this set-up, the total force was a well behaved
function of $\epsilon$, and for every $N$, there was a single point of minimization of the error,
similar to Fig. 4 in \cite{Thompson}, so we set $\epsilon$ as close to this
point as desired.
For the discretization method, since we are using the same $\epsilon$ for every point, 
we looked for placing the points as equally spaced as possible. However, there
is no perfect way to place $N$ equally spaced points on a sphere. Three 
techniques and their implications while using this method were discussed in
\cite{Thompson}. Besides that, the symmetry of the discretization must be taken into consideration.

\begin{figure}%
    \centering
    \subfloat{{\includegraphics[ trim={4.2cm 2.1cm 4.7cm 1.7cm}, clip, width=0.45\linewidth]{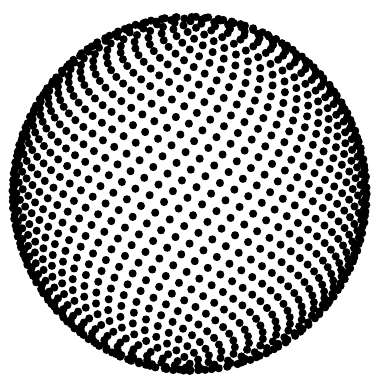}}}
    \qquad
    \subfloat{{\includegraphics[ trim={4.2cm 2.1cm 4.7cm 1.7cm}, clip, width=0.45\linewidth]{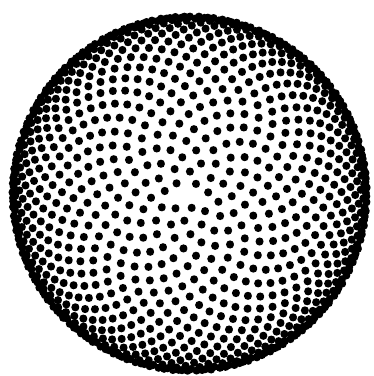}}}
\caption{Side (left) and up view (right) of the discretized sphere with
1800 points and $R=3$, using a Fibonacci lattice.}
    \label{fig:sphere}
\end{figure}

We first tested a Fibonacci lattice since it is a
very simple rule and generates 
very uniform distributions.  The sphere was translated in the $x$-direction. We used $N = 1800$, which showed to be more than
enough for our purposes, and $R=3$ since this is
the radius of the spheres of the swimmer. We compared
the modulus of the total force and torque obtained numerically, which in this case are
respectively:
\begin{equation}
	\bm{F} = \sum^{N}_{k=1} \bm{-f}_k\\
\end{equation}
\begin{equation}
	\bm{T} = \sum^{N}_{k=1} \bm{r}_k \times-\bm{f}_k
\end{equation}
(given the origin set in the central point of the sphere),
with the known analytical expressions for the sphere:
$F =  6 \pi \mu a |u|$ and $T=8\pi\mu a^3|\Omega|$.
For this $N$, we did achieve enough precision for the force
for the value of $\epsilon$ that is shown in Table \ref{tab:1}.
We were getting very proximate values for the total force, however the $y$ and $z$ components
were different from zero by a tiny amount, and we were measuring 
a very small, but not zero net torque, for both
sideways and upward translations. This is  was also reported in \cite{Thompson},
and it is not in agreement with the analytical predictions of zero torque for pure
translations.
This is expected because the discretization is not perfectly symmetric,
as shown in Fig. \ref{fig:sphere}.
However, later we verified that this torque was small enough to be ignored, and 
for this case, we could have just ignored
any rotation or movement out of the $x$ axis.

\begin{figure}
\centering
    \includegraphics[ trim={4.2cm 2.1cm 4.7cm 1.8cm}, clip, scale=1.1]{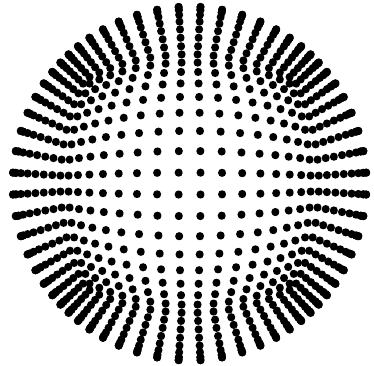}
    \caption{Cubed-sphere discretization for a $16 \times 16$ grid in each face, with $R=3$. This face
   was translated in the $x$ direction.}
\label{fig:cubed_sphere}
\end{figure}

But because of these small discrepancies,
we decided to test another method of discretization, known as cubed-sphere or box to
sphere.
In this discretization, we place the points by projecting a uniform square grid
on the surface of the sphere, as illustrated in Fig. \ref{fig:cubed_sphere}.
Although this discretization is not as uniform as the previous one,
it has multiple planes of symmetry. 
We obtained very precise values for
the total force and, we reproduced exactly the values obtained in \cite{Cortez3d}.
We have now obtained zero torque in every case since  $xy$ and $zx$ are
planes of symmetry.
As it was stated in
\cite{Thompson}, as long as we use a large number of points, the non uniformity
of the discretization is not so important for precision on the total force. However,
we must add that the symmetry may be an important factor, as this case suggests.
For this discretization method, we used grids with $16 \times 16$ points,
that means a total of $N=6 \times16^2 = 1536$ points. The value of $\epsilon$ is shown on Table \ref{tab:1}.
\begin{table}
	\centering
\caption{Values of $N$ and $\epsilon$ used for each discretization type in the simulation of the swimmer.}
\label{tab:1}
\begin{ruledtabular}
\begin{tabular}{lll}
           & Fibonacci lattice & Cubed-sphere \\
\hline
$N$        & 1800              & 1536         \\
$\epsilon$ & 0.0942797519      & 0.1095680485
\end{tabular}
\end{ruledtabular}
\end{table}
\subsection{Simulation of the three-sphere swimmer}%
\label{sub:three_sphere_swimmer}
The swimmer is modelled by three spheres, discretized by the methods discussed
in section \ref{sub:validation_and_tests}. For each discretization type, we used the number of points 
and the values of $\epsilon$ of Table \ref{tab:1}.\\
The method implementation for the swimmer requires some adaptations since now we are 
dealing with moving boundaries. Mainly, we need to recompute matrix $\mathcal{M}$ at each step,
and to determine the velocities of each sphere.
Since we are interested in studying autonomous swimming,
we must find solutions that satisfy at every
step, the following conditions:
\begin{equation}
	\label{eq:forcefree}
	\sum^{N}_{k=1} \bm{f}_{k} = 0
\end{equation}
and:
\begin{equation}
	\label{eq:torquefree}
	\sum^{N}_{k=1} \bm{r}_k \times \bm{f}_k = 0
\end{equation}
which means that the movement does not require any external forces
or torques.
Condition Eq. (\ref{eq:torquefree}) can be satisfied by taking the same precautions as the case
of a single sphere. Again, only by analyzing the swimmer and its symmetry, we
can conclude that no torque should act on it during any of its steps. So if we
use a proper symmetric discretization, this condition is automatically satisfied
in any longitudinal motion of the spheres. However, as in the case
of the Fibonacci lattice, the asymmetry is so small that the resulting small torque
is negligible.

To satisfy Eq. (\ref{eq:forcefree}) we needed a more subtle mechanism. 
\begin{figure}[h]
\centering
    \includegraphics[]{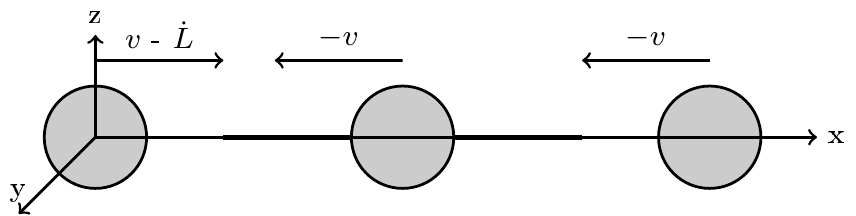}
    \caption{Example of the first step of the swimmer and the
respective velocities.}
\label{fig:esquema}
\end{figure}
We will exemplify how we proceed using the first step as an example, but this
argument is valid for all swimming steps.  By the construction
of the swimmer, at every step, we have the constraint that one arm is
retracting or extending with a given constant rate, which we call $\dot{L}$,
while the other remains fixed.  To satisfy this constraint, 
we can set the velocity of each sphere, as illustrated in Fig. \ref{fig:esquema},
with $\dot{L}$ negative, if the arm is retracting; $v$ is an arbitrary velocity,
and all the vectors are in the direction $\bm{\hat{\imath}}$.
With this setup, for any value of $v$, which is measured relative to the fluid,
we have the execution of step one, but to satisfy Eq. (\ref{eq:forcefree}) we have to find
the specific value of $v$ that will result in a total null force.
Because of the symmetry,  we expect for any motion of this type, that the $y$ and
$z$ force components sum up to zero. If that is the case, the total force will be given
simply by $\bm{F} = F_x \bm{\hat{\imath}}$, and now, because of the linear relation
Eq. (\ref{eq:lin_system}), it will depend linearly on $v$.
\begin{equation}
	F_x = mv + b
	\label{eq:force_lin}
\end{equation}
Using that, we find the correct value of $v$ by
solving two linear systems for the forces with two arbitrary values of $v$,
computing the respective total forces
and with these two values finding the root of Eq. (\ref{eq:force_lin}).
With this $v$ we update the positions with:
\begin{equation}
	\bm{r}_k(t+\delta t) =  \bm{r}_k(t) + \bm{u}_k(t) \delta t
\end{equation}
This process is repeated,
verifying if it is time to go to the next step of the swimming motion
until the cycle is complete. When one complete cycle is executed,
we measure the displacement $\Delta$.
\section{Results}%
\begin{figure}[h]
\centering
    \includegraphics[]{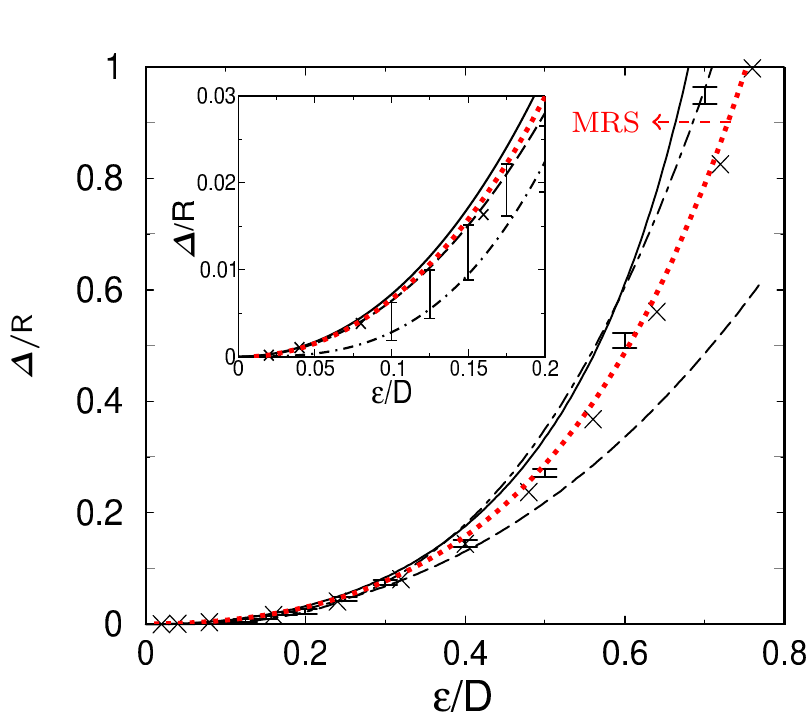}
    \caption{Figure taken from \cite{Yeomans} with our result plotted on top with
	doted line. In this figure, $\bm{\varepsilon}$ is what we denoted as
	$\delta$. The solid line is obtained by solving the Oseen tensor
	equations numerically, the crosses mark is obtained with the lattice
	Boltzmann, the error bars show the results obtained by multiparticle
	collision dynamics, which is a noisy method, the dashed and dotted dashed lines
	are theoretical solutions for $\delta << R$ and $R<<D$ obtained respectively by
Najafi and Golestanian\cite{Najafi} and by the authors of \cite{Yeomans}.}
\label{fig:result}
\end{figure}

\label{sec:results}
Since our aim is to compare the MRS with other methods that were implemented
in \cite{Yeomans}, we used the same parameters of this work: $R=3$ and $D=25$.
The simulation is done by varying the parameter $\delta$ and computing the net displacement
$\Delta$ after one complete cycle.
We present our data in Fig. \ref{fig:result}
by plotting our results directly on top of the data from \cite{Yeomans} (with permission)
\footnote{Reproduced from 
with the permission of AIP Publishing.}.
Our result is shown with a dotted line. The data is presented by
the relation between the dimensionless variables $\Delta/R$ and $\delta/D$. What we call $\delta$
was denoted by $\bm{\varepsilon}$ in the original figure.
We ran two simulations for the swimmer. In each one, we discretized
the spheres by each method discussed previously and
used the same number of points and $\epsilon$ from Table \ref{tab:1}.
However, the results are visually 
indistinguishable, so we are showing only one of the curves.

This figure shows that the results with the MRS
are in very good agreement with the analytical solution (dashed line) for $\delta << D$ and $R << D$.
For higher values of $\delta/D$, when the analytical solutions are no
longer valid, our solutions are very close to the LBM (crosses mark) and MPC
(error bars), both methods that are supposed to work in this range. This
indicates a good behaviour of the MRS for the range of all values of
$\delta/D$.
We note that the dot-dashed line, which is an analytical solution from
\cite{Najafi}, is good for high $\delta/D$ but does not converge for small values,
an assumption initially made in its deduction.
That formula was corrected in \cite{Yeomans} and is shown in Fig. \ref{fig:result} by 
the dashed line.
\section{Conclusion}%
\label{sec:conclusion}
Although the MRS is already being used in a great variety of applications,
we felt that simpler and more careful tests were lacking in the literature, specifically
addressing micro-swimmers, in order to explore the details and capabilities of this method.
Here, we filled this gap by using the MRS to study one of the simplest models of micro-swimmers, the
three-sphere swimmer. This swimmer was already 
studied by other numerical and analytical methods, providing us with material to compare.

First, we have discussed and explained the theory behind the MRS, showing how it is a
different approach to the Stokes equations, and how the regularization of the perturbation
changes the interpretation of the response, increasing the possibilities of use.

We implemented and tested the method for the case of a single translating sphere, showing the importance
of each parameter and discretization type, and how we achieved a balance between precision,
memory usage and speed.
We showed two examples of discretizations and what effects each one had in
the final results, achieving good precision for the total force in both cases.

We then studied the autonomous swim of the three-sphere swimmer numerically.
We modelled the swimmer by using three discretized spheres. We tested both discretization methods,
obtaining similar results for each one.
By comparing our results with results from other methods and with an analytical solution taken
from \cite{Yeomans}, we showed that the MRS performed very well, agreeing 
nicely with the analytical solution in its range of validity, and staying closer
to the LBM results in higher ranges. This is a good indication of the reliability of the method.

We conclude that the MRS is a simple, useful and precise tool to be used in the study of interactions at low Reynolds
number.
\begin{acknowledgements}
I would like to especially thank professors Sandra Prado and Sílvio Dahmen for their helpful suggestions and general guidance during this project.
Also J.M. Yeomans and C.M. Pooley for their willingness to help and provide access to the figure from their paper. This research has been supported by Programa de Iniciação Científica PROPESQ-BIC/UFRGS.
\end{acknowledgements}

\section*{data AVAILABILITY}
The  data  that  support  the  findings  of  this  study  are  available from the corresponding author upon reasonable request.
\bibliography{aipsamp}

\providecommand{\noopsort}[1]{}\providecommand{\singleletter}[1]{#1}
\begin{thebibliography}{16}%
\makeatletter
\providecommand \@ifxundefined [1]{%
 \@ifx{#1\undefined}
}%
\providecommand \@ifnum [1]{%
 \ifnum #1\expandafter \@firstoftwo
 \else \expandafter \@secondoftwo
 \fi
}%
\providecommand \@ifx [1]{%
 \ifx #1\expandafter \@firstoftwo
 \else \expandafter \@secondoftwo
 \fi
}%
\providecommand \natexlab [1]{#1}%
\providecommand \enquote  [1]{``#1''}%
\providecommand \bibnamefont  [1]{#1}%
\providecommand \bibfnamefont [1]{#1}%
\providecommand \citenamefont [1]{#1}%
\providecommand \href@noop [0]{\@secondoftwo}%
\providecommand \href [0]{\begingroup \@sanitize@url \@href}%
\providecommand \@href[1]{\@@startlink{#1}\@@href}%
\providecommand \@@href[1]{\endgroup#1\@@endlink}%
\providecommand \@sanitize@url [0]{\catcode `\\12\catcode `\$12\catcode
  `\&12\catcode `\#12\catcode `\^12\catcode `\_12\catcode `\%12\relax}%
\providecommand \@@startlink[1]{}%
\providecommand \@@endlink[0]{}%
\providecommand \url  [0]{\begingroup\@sanitize@url \@url }%
\providecommand \@url [1]{\endgroup\@href {#1}{\urlprefix }}%
\providecommand \urlprefix  [0]{URL }%
\providecommand \Eprint [0]{\href }%
\providecommand \doibase [0]{http://dx.doi.org/}%
\providecommand \selectlanguage [0]{\@gobble}%
\providecommand \bibinfo  [0]{\@secondoftwo}%
\providecommand \bibfield  [0]{\@secondoftwo}%
\providecommand \translation [1]{[#1]}%
\providecommand \BibitemOpen [0]{}%
\providecommand \bibitemStop [0]{}%
\providecommand \bibitemNoStop [0]{.\EOS\space}%
\providecommand \EOS [0]{\spacefactor3000\relax}%
\providecommand \BibitemShut  [1]{\csname bibitem#1\endcsname}%
\let\auto@bib@innerbib\@empty
\bibitem [{\citenamefont {Gompper}\ \emph {et~al.}(2020)\citenamefont
  {Gompper}, \citenamefont {Winkler}, \citenamefont {Speck}, \citenamefont
  {Solon}, \citenamefont {Nardini}, \citenamefont {Peruani}, \citenamefont
  {Löwen}, \citenamefont {Golestanian}, \citenamefont {Kaupp}, \citenamefont
  {Alvarez}, \citenamefont {Ki{\o}rboe}, \citenamefont {Lauga}, \citenamefont
  {Poon}, \citenamefont {DeSimone}, \citenamefont {Mui{\~{n}}os-Landin},
  \citenamefont {Fischer}, \citenamefont {Söker}, \citenamefont {Cichos},
  \citenamefont {Kapral}, \citenamefont {Gaspard}, \citenamefont {Ripoll},
  \citenamefont {Sagues}, \citenamefont {Doostmohammadi}, \citenamefont
  {Yeomans}, \citenamefont {Aranson}, \citenamefont {Bechinger}, \citenamefont
  {Stark}, \citenamefont {Hemelrijk}, \citenamefont {Nedelec}, \citenamefont
  {Sarkar}, \citenamefont {Aryaksama}, \citenamefont {Lacroix}, \citenamefont
  {Duclos}, \citenamefont {Yashunsky}, \citenamefont {Silberzan}, \citenamefont
  {Arroyo},\ and\ \citenamefont {Kale}}]{Roadmap}%
  \BibitemOpen
  \bibfield  {author} {\bibinfo {author} {\bibfnamefont {G.}~\bibnamefont
  {Gompper}}, \bibinfo {author} {\bibfnamefont {R.~G.}\ \bibnamefont
  {Winkler}}, \bibinfo {author} {\bibfnamefont {T.}~\bibnamefont {Speck}},
  \bibinfo {author} {\bibfnamefont {A.}~\bibnamefont {Solon}}, \bibinfo
  {author} {\bibfnamefont {C.}~\bibnamefont {Nardini}}, \bibinfo {author}
  {\bibfnamefont {F.}~\bibnamefont {Peruani}}, \bibinfo {author} {\bibfnamefont
  {H.}~\bibnamefont {Löwen}}, \bibinfo {author} {\bibfnamefont
  {R.}~\bibnamefont {Golestanian}}, \bibinfo {author} {\bibfnamefont {U.~B.}\
  \bibnamefont {Kaupp}}, \bibinfo {author} {\bibfnamefont {L.}~\bibnamefont
  {Alvarez}}, \bibinfo {author} {\bibfnamefont {T.}~\bibnamefont {Ki{\o}rboe}},
  \bibinfo {author} {\bibfnamefont {E.}~\bibnamefont {Lauga}}, \bibinfo
  {author} {\bibfnamefont {W.~C.~K.}\ \bibnamefont {Poon}}, \bibinfo {author}
  {\bibfnamefont {A.}~\bibnamefont {DeSimone}}, \bibinfo {author}
  {\bibfnamefont {S.}~\bibnamefont {Mui{\~{n}}os-Landin}}, \bibinfo {author}
  {\bibfnamefont {A.}~\bibnamefont {Fischer}}, \bibinfo {author} {\bibfnamefont
  {N.~A.}\ \bibnamefont {Söker}}, \bibinfo {author} {\bibfnamefont
  {F.}~\bibnamefont {Cichos}}, \bibinfo {author} {\bibfnamefont
  {R.}~\bibnamefont {Kapral}}, \bibinfo {author} {\bibfnamefont
  {P.}~\bibnamefont {Gaspard}}, \bibinfo {author} {\bibfnamefont
  {M.}~\bibnamefont {Ripoll}}, \bibinfo {author} {\bibfnamefont
  {F.}~\bibnamefont {Sagues}}, \bibinfo {author} {\bibfnamefont
  {A.}~\bibnamefont {Doostmohammadi}}, \bibinfo {author} {\bibfnamefont
  {J.~M.}\ \bibnamefont {Yeomans}}, \bibinfo {author} {\bibfnamefont {I.~S.}\
  \bibnamefont {Aranson}}, \bibinfo {author} {\bibfnamefont {C.}~\bibnamefont
  {Bechinger}}, \bibinfo {author} {\bibfnamefont {H.}~\bibnamefont {Stark}},
  \bibinfo {author} {\bibfnamefont {C.~K.}\ \bibnamefont {Hemelrijk}}, \bibinfo
  {author} {\bibfnamefont {F.~J.}\ \bibnamefont {Nedelec}}, \bibinfo {author}
  {\bibfnamefont {T.}~\bibnamefont {Sarkar}}, \bibinfo {author} {\bibfnamefont
  {T.}~\bibnamefont {Aryaksama}}, \bibinfo {author} {\bibfnamefont
  {M.}~\bibnamefont {Lacroix}}, \bibinfo {author} {\bibfnamefont
  {G.}~\bibnamefont {Duclos}}, \bibinfo {author} {\bibfnamefont
  {V.}~\bibnamefont {Yashunsky}}, \bibinfo {author} {\bibfnamefont
  {P.}~\bibnamefont {Silberzan}}, \bibinfo {author} {\bibfnamefont
  {M.}~\bibnamefont {Arroyo}}, \ and\ \bibinfo {author} {\bibfnamefont
  {S.}~\bibnamefont {Kale}},\ }\bibfield  {title} {\enquote {\bibinfo {title}
  {The 2020 motile active matter roadmap},}\ }\href {\doibase
  10.1088/1361-648x/ab6348} {\bibfield  {journal} {\bibinfo  {journal} {Journal
  of Physics: Condensed Matter}\ }\textbf {\bibinfo {volume} {32}},\ \bibinfo
  {pages} {193001} (\bibinfo {year} {2020})}\BibitemShut {NoStop}%
\bibitem [{\citenamefont {Childress}(2009)}]{Childress}%
  \BibitemOpen
  \bibfield  {author} {\bibinfo {author} {\bibfnamefont {S.}~\bibnamefont
  {Childress}},\ }\href {\doibase 10.1090/cln/019} {\emph {\bibinfo {title} {An
  Introduction to Theoretical Fluid Mechanics}}}\ (\bibinfo  {publisher} {AMS
  and Courant Institute of Mathematical Sciences at New York University},\
  \bibinfo {year} {2009})\BibitemShut {NoStop}%
\bibitem [{\citenamefont {Fonda}\ and\ \citenamefont
  {Sreenivasan}(2017)}]{Unmixing}%
  \BibitemOpen
  \bibfield  {author} {\bibinfo {author} {\bibfnamefont {E.}~\bibnamefont
  {Fonda}}\ and\ \bibinfo {author} {\bibfnamefont {K.~R.}\ \bibnamefont
  {Sreenivasan}},\ }\bibfield  {title} {\enquote {\bibinfo {title} {{U}nmixing
  demonstration with a twist: {A} photochromic {T}aylor-{C}ouette device},}\
  }\href {\doibase 10.1119/1.4996901} {\bibfield  {journal} {\bibinfo
  {journal} {American Journal of Physics}\ }\textbf {\bibinfo {volume} {85}},\
  \bibinfo {pages} {796--800} (\bibinfo {year} {2017})},\ \Eprint
  {http://arxiv.org/abs/https://doi.org/10.1119/1.4996901}
  {https://doi.org/10.1119/1.4996901} \BibitemShut {NoStop}%
\bibitem [{\citenamefont {Purcell}(1977)}]{Purcell}%
  \BibitemOpen
  \bibfield  {author} {\bibinfo {author} {\bibfnamefont {E.~M.}\ \bibnamefont
  {Purcell}},\ }\bibfield  {title} {\enquote {\bibinfo {title} {{Life at low
  Reynolds number}},}\ }\href {\doibase 10.1119/1.10903} {\bibfield  {journal}
  {\bibinfo  {journal} {American Journal of Physics}\ }\textbf {\bibinfo
  {volume} {45}},\ \bibinfo {pages} {3--11} (\bibinfo {year} {1977})},\ \Eprint
  {http://arxiv.org/abs/https://doi.org/10.1119/1.10903}
  {https://doi.org/10.1119/1.10903} \BibitemShut {NoStop}%
\bibitem [{\citenamefont {Najafi}\ and\ \citenamefont
  {Golestanian}(2004)}]{Najafi}%
  \BibitemOpen
  \bibfield  {author} {\bibinfo {author} {\bibfnamefont {A.}~\bibnamefont
  {Najafi}}\ and\ \bibinfo {author} {\bibfnamefont {R.}~\bibnamefont
  {Golestanian}},\ }\bibfield  {title} {\enquote {\bibinfo {title} {{Simple
  swimmer at low Reynolds number: Three linked spheres}},}\ }\href {\doibase
  10.1103/PhysRevE.69.062901} {\bibfield  {journal} {\bibinfo  {journal} {Phys.
  Rev. E}\ }\textbf {\bibinfo {volume} {69}},\ \bibinfo {pages} {062901}
  (\bibinfo {year} {2004})}\BibitemShut {NoStop}%
\bibitem [{\citenamefont {Golestanian}\ and\ \citenamefont
  {Ajdari}(2008)}]{Najafi2}%
  \BibitemOpen
  \bibfield  {author} {\bibinfo {author} {\bibfnamefont {R.}~\bibnamefont
  {Golestanian}}\ and\ \bibinfo {author} {\bibfnamefont {A.}~\bibnamefont
  {Ajdari}},\ }\bibfield  {title} {\enquote {\bibinfo {title} {{Analytic
  results for the three-sphere swimmer at low Reynolds number}},}\ }\href
  {\doibase 10.1103/PhysRevE.77.036308} {\bibfield  {journal} {\bibinfo
  {journal} {Phys. Rev. E}\ }\textbf {\bibinfo {volume} {77}},\ \bibinfo
  {pages} {036308} (\bibinfo {year} {2008})}\BibitemShut {NoStop}%
\bibitem [{\citenamefont {Nasouri}, \citenamefont {Vilfan},\ and\ \citenamefont
  {Golestanian}(2019)}]{Effic1}%
  \BibitemOpen
  \bibfield  {author} {\bibinfo {author} {\bibfnamefont {B.}~\bibnamefont
  {Nasouri}}, \bibinfo {author} {\bibfnamefont {A.}~\bibnamefont {Vilfan}}, \
  and\ \bibinfo {author} {\bibfnamefont {R.}~\bibnamefont {Golestanian}},\
  }\bibfield  {title} {\enquote {\bibinfo {title} {{Efficiency limits of the
  three-sphere swimmer}},}\ }\href {\doibase 10.1103/PhysRevFluids.4.073101}
  {\bibfield  {journal} {\bibinfo  {journal} {Phys. Rev. Fluids}\ }\textbf
  {\bibinfo {volume} {4}},\ \bibinfo {pages} {073101} (\bibinfo {year}
  {2019})}\BibitemShut {NoStop}%
\bibitem [{\citenamefont {Leoni}\ \emph {et~al.}(2009)\citenamefont {Leoni},
  \citenamefont {Kotar}, \citenamefont {Bassetti}, \citenamefont {Cicuta},\
  and\ \citenamefont {Lagomarsino}}]{Experimental}%
  \BibitemOpen
  \bibfield  {author} {\bibinfo {author} {\bibfnamefont {M.}~\bibnamefont
  {Leoni}}, \bibinfo {author} {\bibfnamefont {J.}~\bibnamefont {Kotar}},
  \bibinfo {author} {\bibfnamefont {B.}~\bibnamefont {Bassetti}}, \bibinfo
  {author} {\bibfnamefont {P.}~\bibnamefont {Cicuta}}, \ and\ \bibinfo {author}
  {\bibfnamefont {M.~C.}\ \bibnamefont {Lagomarsino}},\ }\bibfield  {title}
  {\enquote {\bibinfo {title} {A basic swimmer at low reynolds number},}\
  }\href {\doibase 10.1039/B812393D} {\bibfield  {journal} {\bibinfo  {journal}
  {Soft Matter}\ }\textbf {\bibinfo {volume} {5}},\ \bibinfo {pages} {472--476}
  (\bibinfo {year} {2009})}\BibitemShut {NoStop}%
\bibitem [{\citenamefont {Farzin}, \citenamefont {Ronasi},\ and\ \citenamefont
  {Najafi}(2012)}]{PhysRevE.85.061914}%
  \BibitemOpen
  \bibfield  {author} {\bibinfo {author} {\bibfnamefont {M.}~\bibnamefont
  {Farzin}}, \bibinfo {author} {\bibfnamefont {K.}~\bibnamefont {Ronasi}}, \
  and\ \bibinfo {author} {\bibfnamefont {A.}~\bibnamefont {Najafi}},\
  }\bibfield  {title} {\enquote {\bibinfo {title} {General aspects of
  hydrodynamic interactions between three-sphere low-reynolds-number
  swimmers},}\ }\href {\doibase 10.1103/PhysRevE.85.061914} {\bibfield
  {journal} {\bibinfo  {journal} {Phys. Rev. E}\ }\textbf {\bibinfo {volume}
  {85}},\ \bibinfo {pages} {061914} (\bibinfo {year} {2012})}\BibitemShut
  {NoStop}%
\bibitem [{\citenamefont {Avron}, \citenamefont {Kenneth},\ and\ \citenamefont
  {Oaknin}(2005)}]{Avron_2005}%
  \BibitemOpen
  \bibfield  {author} {\bibinfo {author} {\bibfnamefont {J.~E.}\ \bibnamefont
  {Avron}}, \bibinfo {author} {\bibfnamefont {O.}~\bibnamefont {Kenneth}}, \
  and\ \bibinfo {author} {\bibfnamefont {D.~H.}\ \bibnamefont {Oaknin}},\
  }\bibfield  {title} {\enquote {\bibinfo {title} {Pushmepullyou: an efficient
  micro-swimmer},}\ }\href {\doibase 10.1088/1367-2630/7/1/234} {\bibfield
  {journal} {\bibinfo  {journal} {New Journal of Physics}\ }\textbf {\bibinfo
  {volume} {7}},\ \bibinfo {pages} {234--234} (\bibinfo {year}
  {2005})}\BibitemShut {NoStop}%
\bibitem [{\citenamefont {Earl}\ \emph {et~al.}(2007)\citenamefont {Earl},
  \citenamefont {Pooley}, \citenamefont {Ryder}, \citenamefont {Bredberg},\
  and\ \citenamefont {Yeomans}}]{Yeomans}%
  \BibitemOpen
  \bibfield  {author} {\bibinfo {author} {\bibfnamefont {D.~J.}\ \bibnamefont
  {Earl}}, \bibinfo {author} {\bibfnamefont {C.~M.}\ \bibnamefont {Pooley}},
  \bibinfo {author} {\bibfnamefont {J.~F.}\ \bibnamefont {Ryder}}, \bibinfo
  {author} {\bibfnamefont {I.}~\bibnamefont {Bredberg}}, \ and\ \bibinfo
  {author} {\bibfnamefont {J.~M.}\ \bibnamefont {Yeomans}},\ }\bibfield
  {title} {\enquote {\bibinfo {title} {{Modeling microscopic swimmers at low
  Reynolds number}},}\ }\href {\doibase 10.1063/1.2434160} {\bibfield
  {journal} {\bibinfo  {journal} {The Journal of Chemical Physics}\ }\textbf
  {\bibinfo {volume} {126}},\ \bibinfo {pages} {064703} (\bibinfo {year}
  {2007})},\ \Eprint {http://arxiv.org/abs/https://doi.org/10.1063/1.2434160}
  {https://doi.org/10.1063/1.2434160} \BibitemShut {NoStop}%
\bibitem [{\citenamefont {Cortez}(2001)}]{Cortez2d}%
  \BibitemOpen
  \bibfield  {author} {\bibinfo {author} {\bibfnamefont {R.}~\bibnamefont
  {Cortez}},\ }\bibfield  {title} {\enquote {\bibinfo {title} {{The Method of
  Regularized Stokeslets}},}\ }\href {\doibase 10.1137/S106482750038146X}
  {\bibfield  {journal} {\bibinfo  {journal} {SIAM Journal on Scientific
  Computing}\ }\textbf {\bibinfo {volume} {23}},\ \bibinfo {pages} {1204--1225}
  (\bibinfo {year} {2001})},\ \Eprint
  {http://arxiv.org/abs/https://doi.org/10.1137/S106482750038146X}
  {https://doi.org/10.1137/S106482750038146X} \BibitemShut {NoStop}%
\bibitem [{\citenamefont {Cortez}, \citenamefont {Fauci},\ and\ \citenamefont
  {Medovikov}(2005)}]{Cortez3d}%
  \BibitemOpen
  \bibfield  {author} {\bibinfo {author} {\bibfnamefont {R.}~\bibnamefont
  {Cortez}}, \bibinfo {author} {\bibfnamefont {L.}~\bibnamefont {Fauci}}, \
  and\ \bibinfo {author} {\bibfnamefont {A.}~\bibnamefont {Medovikov}},\
  }\bibfield  {title} {\enquote {\bibinfo {title} {{The method of regularized
  Stokeslets in three dimensions: Analysis, validation, and application to
  helical swimming}},}\ }\href {\doibase 10.1063/1.1830486} {\bibfield
  {journal} {\bibinfo  {journal} {Physics of Fluids}\ }\textbf {\bibinfo
  {volume} {17}},\ \bibinfo {pages} {031504} (\bibinfo {year} {2005})},\
  \Eprint {http://arxiv.org/abs/https://doi.org/10.1063/1.1830486}
  {https://doi.org/10.1063/1.1830486} \BibitemShut {NoStop}%
\bibitem [{\citenamefont {Thompson}(2015)}]{Thompson}%
  \BibitemOpen
  \bibfield  {author} {\bibinfo {author} {\bibfnamefont {T.~B.}\ \bibnamefont
  {Thompson}},\ }\emph {\bibinfo {title} {Exploration of {Local Force
  Calculations Using the Methods of Regularized Stokeslets}}},\ \href@noop {}
  {Master's thesis} (\bibinfo {year} {2015})\BibitemShut {NoStop}%
\bibitem [{\citenamefont {Garcia-Gonzalez}(2016)}]{Jesus}%
  \BibitemOpen
  \bibfield  {author} {\bibinfo {author} {\bibfnamefont {J.}~\bibnamefont
  {Garcia-Gonzalez}},\ }\emph {\bibinfo {title} {{N}umerical analysis of fluid
  motion at low {R}eynolds Number}},\ \href@noop {} {Master's thesis} (\bibinfo
  {year} {2016})\BibitemShut {NoStop}%
\bibitem [{Note1()}]{Note1}%
  \BibitemOpen
  \bibinfo {note} {Reproduced from with the permission of AIP
  Publishing.}\BibitemShut {Stop}%
\end{thebibliography}%

\end{document}